# Electric and Photoelectric Gates for ion backflow suppression in multi-GEM structures

A. Buzulutskov [*], A. Bondar

*Budker Institute of Nuclear Physics, 630090 Novosibirsk, Russia*

**Abstract**

A new approach to suppress ion backflow in multi-GEM structures is suggested. In this approach, the potential difference applied across the gap between two adjacent GEMs is reversed compared to the standard configuration. In such a gap structure, called Electric Gate, a signal transfer from the first to second GEM is presumably provided by the small residual field still existing at small gate voltages and connecting the holes of the two GEMs. On the other hand, ion backflow between the GEMs turned out to be substantially reduced. We also consider another configuration, called Photoelectric Gate, in which in addition to the Electric Gate configuration, a CsI photocathode is deposited on the second GEM. In the Photoelectric Gate, ion backflow through the gap is fully suppressed and the signal transfer through the gap is provided by the photoelectric mechanism due to either avalanche scintillations in the first GEM or proportional scintillations in the electroluminescence gap replacing the first GEM. The idea of the Electric Gate might find applications in the field of TPC detectors and gas photomultipliers. The idea of the Photoelectric Gate is more relevant in the field of two-phase avalanche detectors.

*PACs:* 29.40.C

## 1. Introduction

Ion backflow from the avalanche region to the cathode is one of the central problems in the development of Time Projection Chambers (TPC) and gas photomultipliers. It is desirable to fully suppress ion backflow to prevent field distortion in the TPC drift volume [1] and ion feedback effects in the gas photomultipliers [2]. The ion backflow problem is also important for the performance of two-phase avalanche detectors [3,4]: due to a very low drift velocity in the liquid, positive ions might be accumulated in the liquid layer and screen the electric field.

Using the Gas Electron Multiplier (GEM) [5] as an electron avalanche detector, although does not solve the problem, substantially reduces ion backflow in multi-GEM structures [6,7,8,9,10,11,12], in particular compared to wire chambers. For example, the ion backflow fraction in the triple-GEM detector was measured to be in the range from 2% to 8% at a drift field of 0.5 kV/cm and gain of $10^4$ [8]. The ion backflow fraction is defined as follows:

[*] Corresponding author. Email: buzulu@inp.nsk.su

$$F = (I_C - I_{PI})/I_A \qquad (1),$$

where $I_C$ is the cathode current; $I_A$ the anode current; $I_{PI}$ the primary ionization current in the drift (cathode) gap. Stronger reduction of ion backflow was achieved by using Micro-Hole and Strip Plates (MHSPs) [13]. Nevertheless, the ion backflow reduction obtained is still far from the ultimate suppression, expressed as follows: the ion charge, back-drifting to the cathode from the avalanche region, should be smaller than the primary ionization charge, i.e.

$$F \leq 1/G \qquad (2).$$

Here $G$ is the total gain of the avalanche detector.

The alternative technique to suppress the ion backflow is to operate in a pulsed-gate mode, by applying voltage pulses to wire electrodes [11] or to the GEM itself [14]. The application of this technique is however limited to experiments where the trigger signal can be provided, which is not always possible.

In this paper we consider a new approach to solve the ion backflow problem. In this approach ion backflow is suppressed using the so-called Electric Gate (EG) and Photoelectric Gate (PEG).

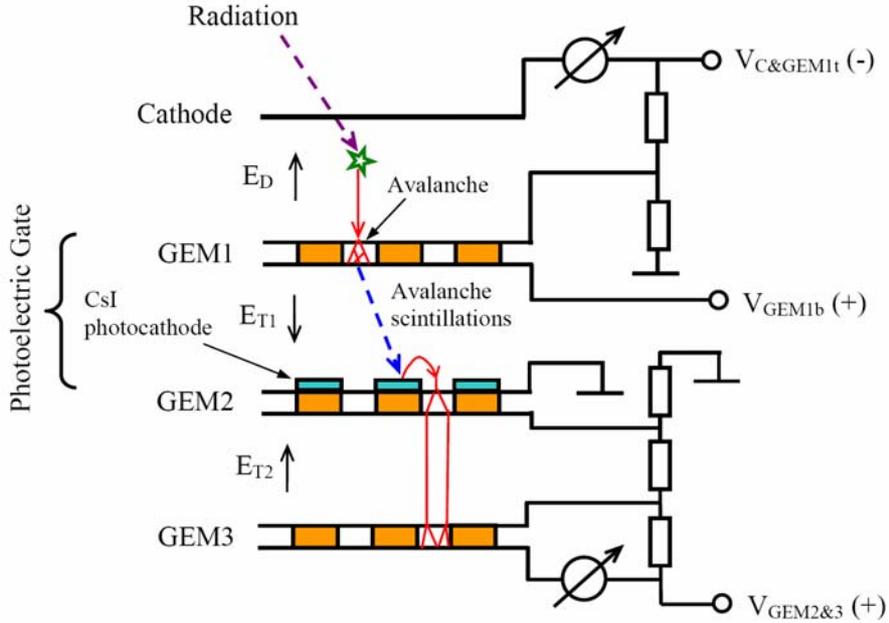

Fig.1 Photoelectric Gate in a triple-GEM detector. (1-2)GEM configuration is depicted.

## 2. Photoelectric Gate

In general, the Electric Gate or Photoelectric Gate can be defined as a structure consisting of an amplification element and a gate gap, the potential difference across the gate gap being reversed compared to the standard configuration. Accordingly, ion backflow through the gap is blocked.

The standard configuration of the multi-GEM detector is defined as that in which the potential differences applied across all GEMs and gaps have the same polarity (see for example [8]). These gaps are the drift gap (between the cathode and the first GEM), the transfer gaps (between the GEMs) and the induction gap (between the GEM and the anode).

The idea of the Photoelectric Gate is explained in Fig.1 by the example of the triple-GEM detector. The transfer field $E_{T1}$ between the first and second GEMs (GEM1 and GEM2 in Fig.1) is reversed compared to the standard configuration, fully blocking ion backflow from the successive GEM elements. In this case the ion backflow to the cathode is derived solely from the first GEM. On the other hand, the signal transfer through the gap is provided by the photoelectric mechanism: avalanche scintillations in the holes of the first GEM induce photoemission from the top electrode of the second GEM coated with a CsI photocathode sensitive in the VUV region [15]. Photoelectrons emitted from the top electrode of the second GEM are effectively collected into its holes and further multiplied if the transfer field $E_{T1}$ is slightly negative, of about 0.1 kV/cm [16].

Apparently, gas mixtures having high scintillation yields in the VUV region should be used. These are known to be noble gases (He, Ne, Ar, Kr and Xe) [17,18] and $CF_4$ [19]. Using noble gases is provided by the unique property of GEM structures to operate in pure noble gases at high gains [20,21,22].

Let us suggest to designate the multi-GEM configuration shown in Fig.1 as the (1-2)GEM mode. Here "minus" indicates that the potential difference between the amplification elements is reversed compared to the standard configuration. The anode or the cathode in this and other configurations is defined according to the standard rule, i.e. as the electrode having the maximum positive or maximum negative potential respectively. According to this definition, the cathode is the first and the anode is the last electrode in Figs. 1, 5, 13, 14.

The standard triple-GEM configuration is designated as (1+2)GEM or just 3GEM. Following Ref. [7], the configuration when the anode signal in the triple-GEM and single-GEM is read out from the additional electrode, i.e. from the printed circuit board (PCB), is denoted as 3GEM+PCB and 1GEM+PCB, respectively. Accordingly, in the 1GEM-PCB mode the induction field $E_I$ (below the GEM) is reversed compared to the standard configuration (see Fig.2).

The principal characteristic of the gate is the gain transfer factor $f$, defined as the transfer efficiency of the signal through the gate gap:

$$G(PEG) = G(GEM1) \cdot f \qquad (3).$$

Here $G(PEG)$ is the gain of the Photoelectric Gate, $G(GEM1)$ the gain of the first GEM. In the Photoelectric Gate this factor can be expressed as follows:

$$f = n_P \cdot \omega_{SA} \cdot \varepsilon_{QE} \cdot \varepsilon_{BS} \cdot \eta_{PE} \qquad (4).$$

Here $n_P$ is the number of scintillation photons emitted per one avalanching electron in the holes of the first GEM; $\omega_{SA}$ the solid angle factor, $\varepsilon_{QE}$ the average quantum efficiency of the CsI photocathode in vacuum in the emission region of the gas; $\varepsilon_{BS}$ the factor accounting for the effect of photoelectron backscattering in gas media; $\eta_{PE}$ the collection efficiency of photoelectrons into the holes of the second GEM.

It should be noted that the gate gain should exceed unity, $G(PEG) \geq 1$, in order to have the detection efficiency of a primary electron (described by Poisson statistics) approaching 100%. Therefore, the gain transfer factor should be as large as possible. Otherwise, one should keep $G(GEM1) \gg 1$, which is not desirable due to enhanced ion backflow from the first GEM.

The gain transfer factor can be estimated experimentally using the gains of the (1-2)GEM and 3GEM configurations, $G((1\text{-}2)GEM)$ and $G(3GEM)$, respectively,

$$f \leq G((1-2)GEM)/G(3GEM) \qquad (5),$$

or even more directly using the gains of the single-GEM configurations

$$f \leq G(1GEM - PCB)/G(1GEM + PCB) \qquad (6)$$

where the PCB is coated with a CsI photocathode. In the latter case, the measurements were carried out using a single-GEM detector shown in Fig. 2.

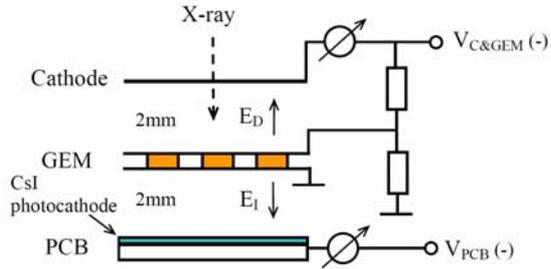

Fig.2 A single-GEM detector to estimate the gain of Photoelectric Gate. 1GEM-PCB configuration is depicted.

The experimental setups in these and other measurements (Figs. 1,2,5) included a high-pressure stainless-steel vessel, inside which a GEM assembly was mounted. The GEM foils were produced by the CERN workshop and had the following parameters: 50 μm thick kapton, 3×3 cm$^2$ active area, 70 and 55 μm hole diameter on the metal and kapton center respectively and 140 μm hole pitch. In the setups of Figs. 1 and 2, a CsI photocathode of a thickness of about 2 μm was deposited on a gold-plated GEM foil using vacuum evaporation equipment. In the setup of Fig. 1, the inter-GEM gaps were 2 mm. In the setup of Fig.2, the gap between the GEM and the PCB was 2 mm. The detectors were operated in a current mode, irradiated with continuous X-rays. The cathode, GEM and PCB electrodes were biased through a resistive high-voltage divider.

The cathode and anode currents, $I_C$ and $I_A$, were measured in the cathode and anode circuit, respectively, as shown in Figs. 1,2 and 5. The gain value is defined as the current of the readout electrode (i.e. as the anode current in Figs. 1 and 5 and as the PCB current in Fig. 2) divided by the primary ionization current in the drift gap, $I_{PI}$. The latter was determined in special measurements, when the drift gap was operated in the ionization mode. Only the currents due to conversion in the drift gap were considered for the gain and ion backflow measurements; the contribution from conversion in successive gaps was measured separately and subtracted.

Fig. 3 shows the gain in the 1GEM±PCB modes of operation of the single-GEM detector as a function of the PCB voltage at a fixed GEM voltage, in Kr and Ar+10%CF$_4$. In Kr, the photoelectric signal was observed in the 1GEM-PCB mode. Its value at $V_{PCB}$=-300V, corresponding to the induction field of -1.5 kV/cm, was by a factor of 50-100 smaller compared to the 1GEM+PCB mode (at $V_{PCB}$=+300V), indicating that the gain transfer factor is rather small. At smaller induction fields the photoelectric signal is also smaller, obviously due to the effect of photoelectron backscattering which is particularly strong in noble gases [23,24].

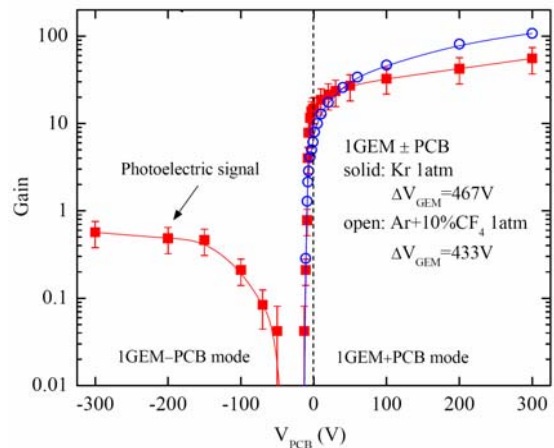

Fig.3 Gain in the 1GEM±PCB modes of operation of the single-GEM detector of Fig. 2, as a function of the PCB voltage, in Kr and Ar+10%CF$_4$ at 1 atm, at a fixed GEM voltage. The PCB is coated with a CsI photocathode.

In Ar+10%CF$_4$, the photoelectric signal was at least by two orders of magnitude weaker than that in Kr, indicating that the gain transfer factor in this mixture is much smaller than that in Kr, apparently due to the lower scintillation yield in CF$_4$ compared to noble gases [19].

Fig. 4 shows the gain in the 1GEM±PCB modes as a function of the voltage applied across the GEM (GEM voltage, $\Delta V_{GEM}$) at a fixed PCB voltage in Kr and Ar+10%CF$_4$. In Kr the gain in the 1GEM-PCB (photoelectric) mode is by a factor of 40-60 lower than that in the 1GEM+PCB mode, again indicating that the gain transfer factor is small: $f\sim1/40$-$1/60$.

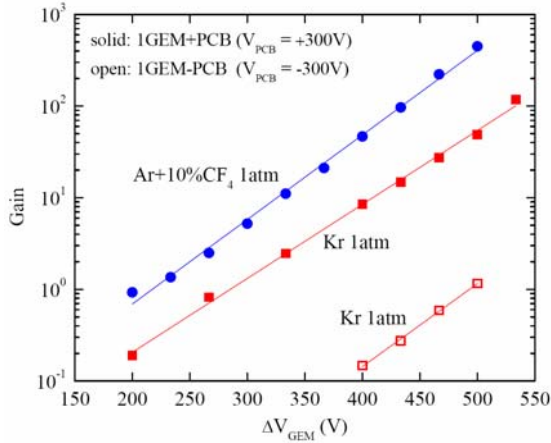

Fig.4 Gain in the 1GEM±PCB modes of operation of the single-GEM detector of Fig. 2, as a function of the GEM voltage at a fixed PCB voltage, of ±300 V respectively, in Kr and Ar+10%CF$_4$ at 1 atm. The PCB is coated with a CsI photocathode.

We also tested the triple-GEM detector with the Photoelectric Gate shown in Fig. 1, i.e. operated in the (1±2)GEM modes, in He at 5 atm. Similarly to the previous measurements, the gain transfer factor in the (1-2)GEM (photoelectric) mode turned out to be small: of about 1/30.

Such a small value of the gain transfer factor is not surprising. Indeed, if realistic values of the terms in expression (4) are taken, $\varepsilon_{QE}=0.2$, $n_P=1$, $\omega_{SA}=1/3$ and $\varepsilon_{BS}=1/3$, the transfer factor would be $f\sim1/50$, which is close to the observed value. Since the value of $n_P$ for noble gases is generally unknown, we took here the largest value ever reported for other gases. Obviously, the low efficiency of the signal transfer in the Photoelectric Gate does not allow one to reach the ultimate suppression of ion backflow (2) due to the large contribution from the first GEM.

## 3. Electric Gate

Looking carefully at Fig. 3, one can observe that the avalanche (electric) signal still exists when the PCB voltage becomes negative and small in magnitude. We studied this effect using the triple-GEM detector and the structure called Electric Gate shown in Fig. 5.

In contrast to the Photoelectric Gate, the CsI photocathode is not used in the Electric Gate and the contribution of the photoelectric signal is not important. Similarly to the Photoelectric Gate, the potential difference across the gate gap is reversed compared to the standard configuration, but its magnitude is rather small, of the order of -10 V. In these measurements, the gate gap was 1.6 mm and the gas filling was either He at 5 atm, of a purity of 99.99%, or Ar+10%CF$_4$ at 1 atm. The experimental procedure was similar to that described in the previous section.

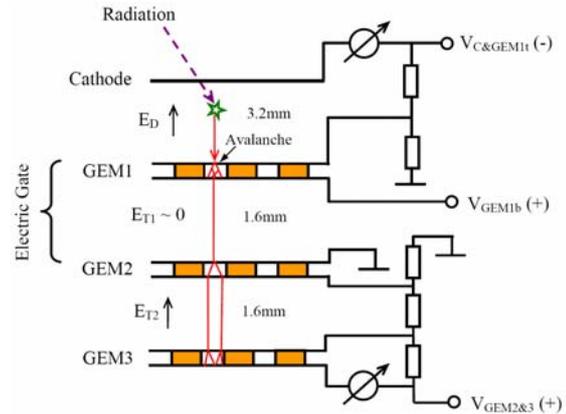

Fig.5 Electric Gate in a triple-GEM detector. (1-2)GEM configuration is depicted.

Fig. 6 shows the anode and cathode currents normalized to the primary ionization current (i.e. exactly the gain in the case of the anode current) and the ion backflow fraction in the (1±2)GEM modes of operation of the triple-GEM detector as a function of the voltage applied across the gate gap. The data were obtained at fixed GEM voltages, in He at 5 atm.

The gate voltage is designated as $\Delta V_{EG}$ and is equal to -$V_{GEM1b}$ in Fig. 5. In this notation, the gate is open when the gate voltage is positive ($\Delta V_{EG}>0$), corresponding to operation in the (1+2)GEM mode, and the gate is close when the gate voltage is negative ($\Delta V_{EG}<0$), corresponding to operation in the (1-2)GEM mode.

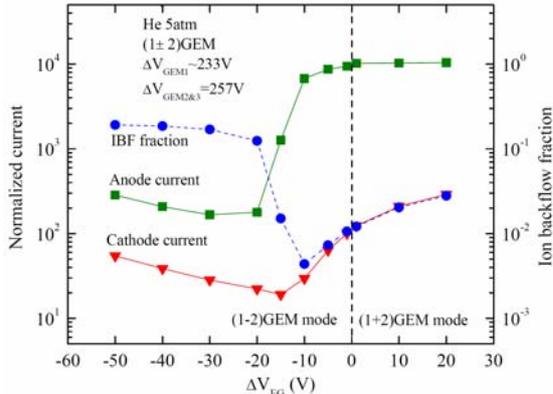

Fig.6 Anode and cathode currents normalized to the primary ionization current (left scale) and ion backflow (IBF) fraction (right scale) in the (1±2)GEM modes of operation of the triple-GEM detector of Fig.5, as a function of the voltage across the gate gap at fixed GEM voltages, in He at 5 atm.

One can see that at $\Delta V_{EG}<0$, i.e. in the (1-2)GEM mode, the cathode current decreases faster than the anode current when decreasing the gate voltage. Accordingly, the ion backflow substantially decreases and reaches a minimum at $\Delta V_{EG}$=-10 V: here ion backflow is by almost an order of magnitude smaller than that in the (1+2)GEM mode. At the same time, the gain transfer factor here is still large enough. This means that the Electric Gate, operated at small voltages, may help to substantially reduce ion backflow.

The transfer of the electron signal from the first to second GEM in the (1-2)GEM mode is most probably provided by the following fact: when the gate voltage is reversed but small in magnitude, there exist a residual field connecting the holes of the adjacent GEMs, along which electrons and ions can drift. This is clear from Fig. 7, showing drift paths of electrons generated uniformly across the hole of the first GEM, in the (1-2)GEM mode at a gate voltage of -10 V (diffusion is disregarded). This plot was obtained using MAXWELL [25] and GARFIELD [26] simulation programs [27].

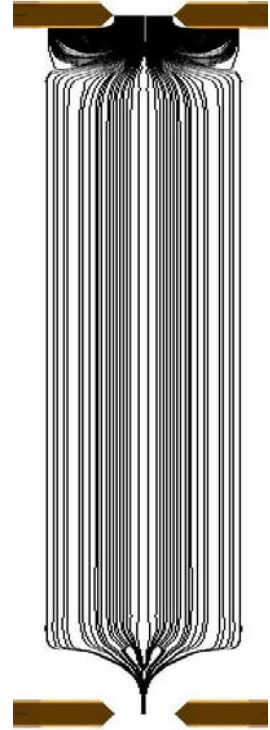

Fig.7 Drift paths of electrons in the Electric Gate shown in Fig. 5, in the (1-2)GEM mode in Ar+10%$CF_4$ at 1 atm. The gate voltage is $\Delta V_{EG}$ =-$V_{GEM1b}$=-10 V; the GEM voltages are $\Delta V_{GEM1}$=$\Delta V_{GEM2}$=400 V. Diffusion is disregarded. The vertical scale is reduced by a factor of 4 compared to the horizontal scale.

One can see that the polarity of the gate field does not always coincide with the polarity of the gate voltage: while the gate voltage is reversed in the (1-2)GEM mode, the residual field is not reversed, unless the gate voltage is below -20 V (see the next paragraph). For this reason, the Electric Gate in the (1-2)GEM mode is not fully closed at small gate voltages: instead, it is partially open both for electrons and ions. But for ions it is less open, as observed in the experiment, resulting in the substantial reduction of ion backflow. Quite possible this effect exploits the large difference in diffusion between electrons and ions. One can also see from Fig. 7 that just the central regions of the GEM holes, in which the electric field is the highest, are connected by field lines. This explains why the gain transfer factor can be relatively large in the Electric Gate: electron avalanches tend to develop just in the high field regions.

Simulations showed that at gate voltages below -20 V the gate is fully closed: there are no field lines connecting the holes of the adjacent GEMs any more. At these voltages the anode signal abruptly drops, presumably indicating a change of the field polarity. It is interesting however that the anode signal still exists (see Fig. 6). Most probably, the gate operates here in the photoelectric mode, like that shown in Fig. 1, provided by the fact that the quantum efficiency of the copper electrode of the second GEM is high enough in the region of He scintillations. The alternative explanation that the anode signal here is provided by electron diffusion across the gap, against the electric field, is ruled out by calculations.

Fig. 8 compares the gain characteristics, Figs. 9 and 10 the ion backflow characteristics and Fig. 11 the anode signals in the (1-2)GEM and 3GEM mode of operation of the triple-GEM detectors, in He and Ar+10%CF$_4$. In these figures, the gate voltage in the (1-2)GEM mode was proportional to the GEM voltage: $\Delta V_{EG}$=-0.038$\Delta V_{GEM}$. This corresponds to gate voltages of about -10 V in He and -15 V in Ar+10%CF$_4$. In the 3GEM mode, the voltages across each GEM and each inter-GEM gap were equal. In both mixtures the gain transfer factor, estimated from expression (5), is of the order of 1/10 at gains of $10^4$ (see Fig. 8), which is substantially larger than that in the Photoelectric Gate.

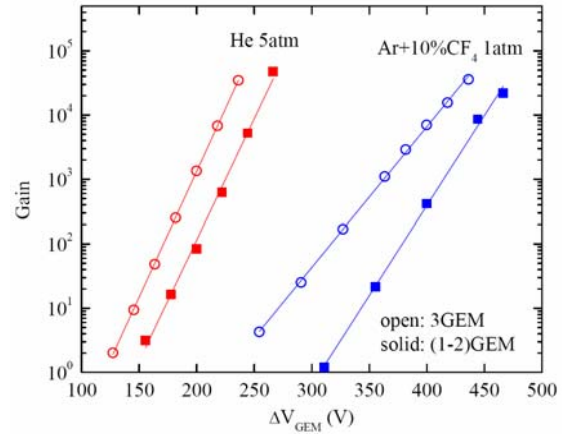

Fig.8 Gain in the (1-2)GEM and 3GEM mode of operation of the triple-GEM detector of Fig. 5, as a function of the GEM voltage, in He at 5 atm and Ar+10%CF$_4$ at 1 atm. In the (1-2)GEM mode, the gate voltage is proportional to the GEM voltage: $\Delta V_{EG}$=-0.038$\Delta V_{GEM}$.

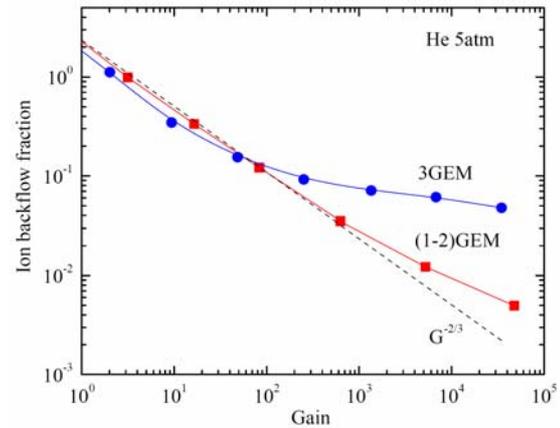

Fig.9 Ion backflow fraction as a function of the gain in the (1-2)GEM and 3GEM mode of operation of the triple-GEM detector of Fig. 5, in He at 5 atm. In the (1-2)GEM mode, the gate voltage is proportional to the GEM voltage: $\Delta V_{EG}$=-0.038$\Delta V_{GEM}$. The drift field in the cathode gap is proportional to the GEM voltage and at a gain of $10^4$ is equal to 0.39 and 0.35 kV/cm in the (1-2)GEM and 3GEM mode respectively. The dashed line shows the $G^{-2/3}$ dependence.

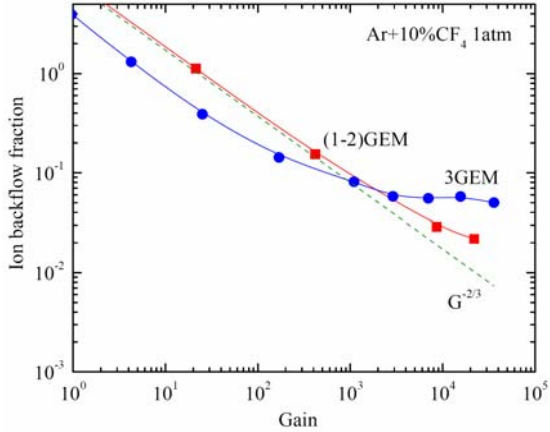

Fig.10 Ion backflow fraction as a function of the gain in the (1-2)GEM and 3GEM mode of operation of the triple-GEM detector of Fig. 5, in Ar+10%CF$_4$ at 1 atm. In the (1-2)GEM mode, the gate voltage is proportional to the GEM voltage: $\Delta V_{EG}$=-0.038$\Delta V_{GEM}$. The drift field in the cathode gap is proportional to the GEM voltage and at a gain of $10^4$ is equal to 0.70 and 0.64 kV/cm in the (1-2)GEM and 3GEM mode respectively. The dashed line shows the $G^{-2/3}$ dependence.

The effect of the Electric Gate on reducing ion backflow is distinctly seen in both mixtures: at a gain of $10^4$ the ion backflow fraction in the (1-2)GEM mode is reduced by a factor 6 and 2 compared to that in the 3GEM mode, in He and Ar+10%CF$_4$, respectively (Figs. 9 and 10).

The small value of the transfer field in the gate gap in the (1-2)GEM mode may result in increasing the signal length. This was in particular observed in He: the rise-time of the anode pulse in Fig. 11, corresponding to the duration of the avalanche signal, is about 3.2 µs in the (1-2)GEM mode, which is by a factor of 4 larger that that in the 3GEM mode.

It is interesting that the ion backflow fraction depends on the gain as $G^{-2/3}$ (here $G$ is the total gain) in a wide gain range (Figs. 9 and 10). This indicates that the major contribution to ion backflow is provided by the first GEM, since it was shown [8] that

$$F = aG^{-2/3} + bG^{-1/3} + c \quad (7),$$

where the first, second and third terms describe the contribution of the first, second and third GEM, respectively. Therefore, it is desirable to have as low gain of the first GEM as possible, in order to further reduce ion backflow.

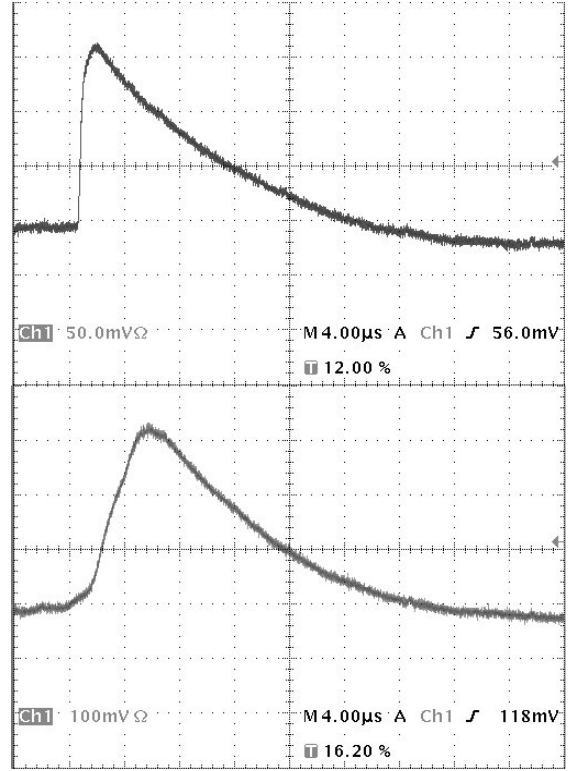

Fig.11 Typical anode signals in the (1-2)GEM (bottom) and 3GEM (top) mode of operation of the triple-GEM detector of Fig. 5, detected with a charge-sensitive amplifier, induced by X-rays in He at 5 atm, at a gain of $3.5\times10^4$ and $4.8\times10^4$ respectively. In the (1-2)GEM mode, the gate voltage is proportional to the GEM voltage: $\Delta V_{EG}$=-0.038$\Delta V_{GEM}$.

This was realized by operating the triple-GEM detector in the (1-2)GEM mode at a fixed voltage of the first GEM, in He (see Fig. 12). The gate voltage was also fixed and was equal to -10 V. The gate gain $G(EG)$ was determined experimentally using the expression

$G(EG) = G((1-2)GEM)/G(2GEM)$ (8),

where $G(2GEM)$ is the gain of the double-GEM structure consisting of the second and third GEMs. An excellent result was obtained as one can see from Fig. 12. At a gate gain of 3 the ion backflow fraction in the (1-2)GEM mode reaches a rather small value: of about $3\times10^{-3}$ at a gain of $10^4$ and drift field of 0.4 kV/cm. This is about a factor of 20 smaller compared to the 3GEM mode. Now the ion backflow fraction tends to depend on the gain as $G^{-1}$, since it is mostly determined by the constant contribution from the first GEM.

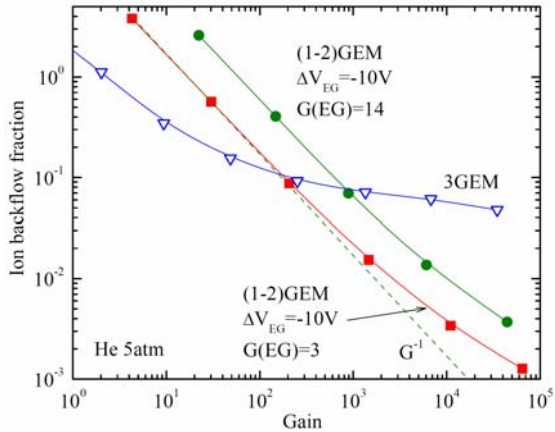

Fig.12 Ion backflow fraction as a function of the gain in the (1-2)GEM and 3GEM mode of operation of the triple-GEM detector of Fig. 5, in He at 5 atm. In the (1-2)GEM mode, the gate voltage is constant $\Delta V_{EG}$=-10 V. The voltage of the first GEM is also constant, having certain values corresponding to the gate gain $G(EG)$=3 and $G(EG)$=14; the drift field in the cathode gap is also constant and equal to 0.36 and 0.42 kV/cm, respectively. The dashed line shows the $G^{-1}$ dependence.

## 4. Photoelectric Gate with electroluminescence gap

It was shown in section 2 that the performance of the Photoelectric Gate is not very efficient in terms of the ion backflow suppression due to two problems. These are the small value of the gain transfer factor and the large ion backflow contribution from the first amplification element.

In this section we discuss possible solutions of these problems (not yet tested experimentally). We suggest to replace the first GEM by a dedicated electroluminescence gap formed by two wire meshes (see Fig. 13). In this structure, ion backflow from the first element is fully eliminated since the electrons are not multiplied in the gap any more. The photoelectric mechanism of the gain transfer is now provided by proportional scintillations in the gap under moderate electric field. For this, heavy noble gases should be used, namely Ar, Kr and Xe, since these are known to effectively produce proportional scintillations [28].

The gap should be thick enough to produce enough light. For example, a 1 cm thick electroluminescence gap can produce as much as 200 VUV photons per electron in Xe at atmospheric pressure, at a field of 4 kV/cm [28,29]. Such amount of photons would be enough for efficient signal transfer through the gate: according to (4) the gain transfer factor would exceed unity in this case. In addition, the scintillation yield of the gate can be further increased by increasing the gas pressure and the gap voltage. It should be noted that the role of the electroluminescence gap at higher pressures could presumably be played by the so-called Thick GEM [30] operated in a proportional scintillation mode.

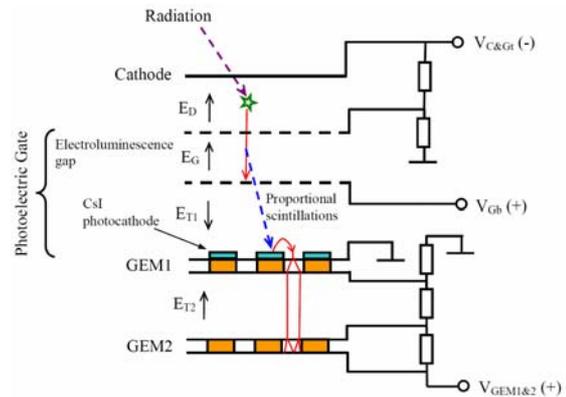

Fig.13 Photoelectric Gate with electroluminescence gap in a double-GEM detector.

Apparently, the Photoelectric Gate with the electroluminescence gap would satisfy the condition of ultimate suppression of ion backflow (2). However, its spatial resolution (of the order of cm) and timing resolution will be limited due to the large thickness of the electroluminescence gap. The solution of this problem is suggested in Fig. 14 using the concept of the Two-Phase Photoelectric Gate relevant to two-phase avalanche detectors [3,4]: the electroluminescence gap is placed in the noble liquid, while the multi-GEM multiplier is kept in the gas phase. Since the noble liquid density is by a factor of 500-800 higher than that of the noble gas, the thickness of the electroluminescence gap can be as low as 50 μm. Therefore, its role might be played by the electroluminescence plate made of a GEM- or MHSP-like structure operated in a proportional scintillation mode. The electric field in the gate gap, between the electroluminescence plate and the first GEM, is reversed compared to the standard configuration, fully blocking ion backflow through the gap. Note that the use of the Two-Phase Photoelectric Gate in two-phase avalanche detectors allows one to avoid electron emission through the liquid-gas interface, which might be very useful for stable operation at high gains.

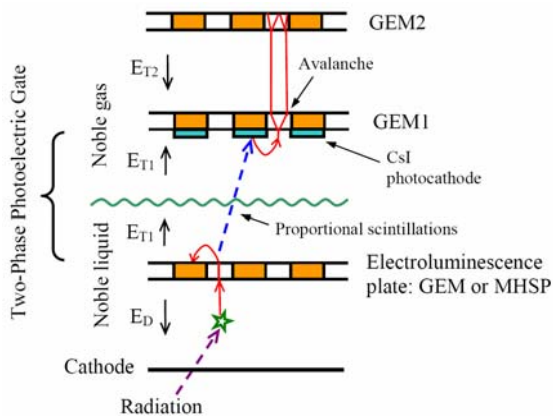

Fig.14 Two-Phase Photoelectric Gate with electroluminescence plate in a two-phase double-GEM detector.

It should be noted that proportional scintillations and electron avalanching in liquid Xe were successfully observed earlier in proportional counters [31] and micro-strip plates [32]. Also, earlier works on the hybrid gas detectors, recording avalanche and proportional scintillations using gas detectors sensitive in the VUV region [33,34,35], might be useful for the development of the Photoelectric Gate.

## 5. Summary

In this paper we suggested a new approach to suppress ion backflow in multi-GEM structures. In this approach, the potential difference applied across the gap between two adjacent GEMs is reversed compared to the standard configuration. Such a gap structure is called Electric Gate or Photoelectric Gate.

In the Electric Gate, the signal transfer from the first to second GEM is presumably provided by the small residual field, still existing at small gate voltages and connecting the holes of the two GEMs. On the other hand, ion backflow between the GEMs turned out to be substantially reduced: the ion backflow fraction in the triple-GEM detector having the Electric Gate was measured to be as low as $3\times10^{-3}$ at a gain of $10^4$ and drift field of 0.4 kV/cm. This is by a factor of 20 lower than that in the standard triple-GEM configuration. The optimization of the gate parameters, in particular the gap thickness, GEM and gate voltages, GEM hole pitch and gas composition, may help to further reduce ion backflow.

We also presented the idea of the Photoelectric Gate, in which in addition to the Electric Gate configuration, a CsI photocathode is deposited on the second GEM. In the Photoelectric Gate, ion backflow through the gap is fully suppressed and the signal transfer through the gap is provided by photoelectric mechanism due to either avalanche scintillations in the holes of the first GEM or proportional scintillations in a dedicated electroluminescence gap replacing the first GEM. The modification of the latter structure is the Two-Phase Photoelectric Gate, where the role of the electroluminescence gap is played by a thin electroluminescence plate placed in the liquid; it is relevant to two-phase avalanche detectors. At the moment, the efficiency of the signal transfer through the Photoelectric Gate consisting of two GEM elements was measured to be rather low, of the order of 1/50 in Kr. This is not enough for ultimate suppression of ion backflow due to the large contribution from the first GEM. In the Photoelectric Gate with the electroluminescence gap this problem will hopefully be solved.

It should be noted that the idea similar to that of the Photoelectric Gate has been independently suggested by other authors [36]: they suggested however to use MHSPs instead of GEMs, which in principle might further reduce ion backflow from the first amplification element and increase the gate gain due to the solid angle effect.

The idea of the Electric Gate might find applications in the field of TPC detectors and gas photomultipliers. The idea of the Photoelectric Gate is more relevant in the field of two-phase avalanche detectors.

## Acknowledgements

We thank A. Vasiljev for computing electric field maps in GEM structures. The research described in this publication was made possible in part by Award 04-78-6744 of INTAS Grant and in part in the frame of the ILC TPC collaboration.

## References

1. TESLA Technical Design Report, Part IV, A Detector for TESLA, Eds. T. Behnke, S. Bertolucci, R.D. Heuer, R. Settles, DESY 2001-011, ECFA 2001-209, (2001).
2. A. Breskin, D. Mormann, A. Lyashenko, R. Chechik, F.D. Amaro, J.M. Maia, J.F.C.A. Veloso, J.M.F. dos Santos, Nucl. Instr. and Meth. A 553 (2005) 46, and references therein.
3. A. Buzulutskov, A. Bondar, L. Shekhtman, R. Snopkov, Y. Tikhonov, IEEE Trans. Nucl. Sci 50 (2003) 2491.
4. A. Bondar, A. Buzulutskov, A. Grebenuk, D. Pavlyuchenko, R. Snopkov, Y. Tikhonov, Nucl. Instr. and Meth. A 556 (2006) 273, and references therein.
5. F. Sauli, Nucl. Instr. and Meth. A 386 (1997) 531.
6. S. Bachmann, A. Bressan, L. Ropelewski, F. Sauli, A. Sharma, D. Mormann, Nucl. Instr. and Meth. A 438 (1999) 376.
7. A. Breskin, A. Buzulutskov, R. Chechik, B.K. Singh, A. Bondar, L. Shekhtman, Nucl. Instr. and Meth. A 478 (2002) 225.
8. A. Bondar, A. Buzulutskov, L. Shekhtman, A. Vasiljev, Nucl. Instr. and Meth. A 496 (2003) 325.
9. F. Sauli, S. Kappler, L. Ropelewski, IEEE Trans. Nucl. Sci. 50 (2003) 803.
10. S. H. Park et al., J. of Korean Phys. Soc. 43 (2003) 332.
11. D. Mormann, A. Breskin, R. Chechik, D. Bloch, Nucl. Instr. and Meth. A 516 (2004) 315.
12. M. Killenberg, S. Lotze, J. Mnich, A. Münnich, S. Roth, F. Sefklow, M. Tonutti, M. Weber, P. Wienemann, Nucl. Instr. and Meth. A 530 (2004) 251.
13. J.F.C.A. Veloso, F.D. Amaro, J.M. Maia, A.V. Lyashenko, A. Breskin, R. Chechik, J.M.F. dos Santos, O. Bouianov, M. Bouianov, Nucl. Instr. and Meth. A 548 (2005) 375, and references therein.
14. F. Sauli, L. Ropelewski, P. Everaerts, Nucl. Instr. and Meth. A 560 (2006) 269.
15. A. Breskin, Nucl. Instr. and Meth. A 371 (1996) 116, and references therein.
16. D. Mormann, A. Breskin, R. Chechik, C. Shalem, Nucl. Instr. and Meth. A 530 (2004) 258.
17. S. Kubota, T. Takahashi, T. Doke, Phys. Rev 165 (1968) 225.
18. M. Suzuki, S. Kubota, Nucl. Instr. and Meth. 164 (1979) 197.
19. A. Pansky, A. Breskin, A. Buzulutskov, R. Chechik, V. Elkind, J. Vavra, Nucl. Instr. and Meth. A 354 (1995) 262.
20. A. Buzulutskov, A. Breskin, R. Chechik, G. Garty, F. Sauli, L. Shekhtman, Nucl. Instr. and Meth. A 443 (2000) 164.
21. A. Buzulutskov, Nucl. Instr. and Meth. A 494 (2002) 148, and references therein.
22. A. Buzulutskov, J. Dodd. R. Galea, Y. Ju, M. Leltchouk, P. Rehak, V. Tcherniatine, W. J. Willis, A. Bondar, D. Pavlyuchenko, R. Snopkov, Y. Tikhonov, Nucl. Instr. and Meth. A 548 (2005) 487.
23. A. Breskin, A. Buzulutskov, R. Chechik, A. Di Mauro, E. Nappi, G. Paic, F. Piuz, Nucl. Instr. and Meth. A 367 (1995) 342.
24. T.H.V.T. Dias, P.J.B.M. Rachinhas, J.A.M. Lopes, F.P. Santos, L.M.N. Tavora, C.A.N. Conde, A.D. Stauffer, J. Phys. D 37 (2004) 540.
25. MAXWELL: a program for 3-D field calculations, see http://wwwce.web.cern.ch/wwwce/ae/Maxwell/.
26. GARFIELD: a program to simulate gas detectors, see http://consult.cern.ch/writeup/garfield/.
27. A. Vasiljev, private communication.
28. A.J.P.L. Policarpo, Nucl. Instr. and Meth. A 196 (1982) 53.
29. A. Bolozdynya, Nucl. Instr. and Meth. A 422 (1999) 314.

30. C. Shalem, R. Chechik, A. Breskin, K. Michaeli, Nucl. Instr. and Meth. A 558 (2006) 475.
31. K. Masuda et al., Nucl. Instr. and Meth. 160 (1979) 247.
32. A.J.P.L. Policarpo et al., Nucl. Instr. and Meth. A 365 (1995) 568.
33. P. Fonte, V. Peskov, F. Sauli, Nucl. Instr. and Meth. A 310 (1991) 140.
34. H. Brauning et al., Nucl. Instrum. and Meth. A 348 (1994) 223.
35. C.M.B. Monteiro et al., Nucl. Instr. and Meth. A 490 (2002) 169.
36. J.F.C.A. Veloso, F.D. Amaro, J.M.F. dos Santos, A. Breskin, A. Lyashenko, R. Chechik, The Photon-Assisted Cascaded Electron Multiplier: a Concept for Potential Avalanche-Ion Blocking, Eprint http://arxiv.org/physics/0606209, submitted to JINST.